\documentclass[runningheads]{llncs}
\usepackage{enumitem}
\usepackage{graphicx}
\usepackage[portuguese]{babel}
\usepackage{listings}
\usepackage{fancyvrb}
\fvset{%
fontsize=\small,
numbers=left
}

\begin{document}
\pagenumbering{gobble}
\pagestyle{plain}

\title{JepREST: Teste Funcional de Aplicações REST Distribuídas}
%
%
\author{Sara Simões, Ana Ribeiro, Carla Ferreira e Nuno Preguiça}
\authorrunning{S. Simões, A. Ribeiro, C. Ferreira, N. Preguiça}
\institute{Faculdade de Ciências e Tecnologia da Universidade Nova de Lisboa}

\maketitle  
%
\begin{abstract}
As aplicações móveis e Web são frequentemente suportadas
por serviços aplicacionais com interface REST, 
implementados usando um conjunto de componentes 
distribuídos que interagem entre si. 
Esta aproximação permite que os serviços apresentem 
elevada disponibilidade e desempenho, a um custo inferior 
ao de um sistema monolítico. 
Porém, a existência de múltiplos componentes torna o 
processo de desenvolvimento destes sistemas mais complexo 
e por isso, suscetível à existência de erros.
Neste trabalho apresentamos o JepREST, um sistema
que permite automatizar a utilização das bibliotecas
Jepsen para testar a correção de aplicações distribuídas
que fornecem uma interface REST. 
A partir de uma especificação da interface do 
serviço, o JepREST gera e executa um conjunto de 
testes com múltiplos clientes a efetuarem operações 
concorrentemente, verificando posteriormente se o 
comportamento do sistema é linearizável. 
A avaliação preliminar efetuada mostra que o
JepREST permite simplificar o teste de aplicações REST. 
\end{abstract}
\section{Introdução}

As aplicações móveis e Web são frequentemente implementadas utilizando 
serviços aplicacionais de suporte que executam a lógica da aplicação e mantêm o seu estado, 
fornecendo uma interface REST usadas pelos clientes para executar operações na aplicação.
Em aplicações não triviais, estes serviços 
aplicacionais 
são sistemas distribuídos complexos,
compostos por vários componentes, incluindo serviço de \textit{cache} aplicacionais (e.g., Memcached, Redis),
serviços de disseminação de eventos (e.g. Kafka), bases de dados e serviços de armazenamento 
de dados replicados.

A complexidade destes sistemas advém do objetivo de permitir um elevado desempenho, disponibilidade e
tolerância a falhas. Em consequência, estes sistemas tornam-se também difíceis de desenhar e programar,
sendo suscetíveis à ocorrência de erros. 
Assim, a validação e teste de sistemas aplicacionais é
um problema importante e o teste de aplicações distribuídas complexas tem atraído a atenção 
da comunidade de investigação e da indústria~\cite{netflix,jepsenElleGitHub,Alvaro17Abstracting}. 

Existem várias ferramentas para testar serviços REST~\cite{load_testing_tools}, em particular ferramentas que permitem executar testes de carga, avaliando o desempenho dos serviços REST. 
Algumas destas ferramentas (e.g. Artillery~\cite{artillery-docs}) permitem efetuar testes funcionais, nos quais se verifica se os serviços se comportam de acordo com o esperado. 
No entanto, o suporte para efetuar testes funcionais é limitado e complexo. Primeiro, deve ser o utilizador da ferramenta a especificar o resultado esperado da execução de uma operação. Se é possível fazê-lo em testes simples, torna-se  difícil ou mesmo impossível fazê-lo em testes aleatórios e que envolvam a execução concorrente de operações. 
Segundo, estas ferramentas não têm mecanismos para perturbar a execução dos componentes do serviço, testando o comportamento do sistema em situações em que ocorram falhas. 
Assim, tem sido realçada a necessidade de criar aplicações que simplifiquem a criação de testes de aplicações distribuídas~\cite{Alvaro17Abstracting}.

Este artigo apresenta o JepREST, um sistema para simplificar 
o teste de aplicações distribuídas com interfaces REST, cujo objetivo é analisar a correção de uma aplicação REST quando a mesma é submetida a um conjunto de testes funcionais, onde existem múltiplos clientes a executarem pedidos concorrentes, podendo ocorrer falhas nos vários componentes da aplicação em estudo.

O JepREST é constituído por quatro componentes.
O primeiro componente gera, a partir da especificação do serviço REST, 
o código para invocar operações no serviço REST. 
O segundo componente gera o código para a execução de cargas
de trabalhos definidas pelo programador, incluindo o código para a criação de novos objetos na aplicação. 
O terceiro componente executa os testes, controlando a execução da 
aplicação REST e dos clientes de teste do serviço. 
O último componente verifica se a execução corresponde a um comportamento
correto da aplicação REST, definido como uma execução que seja linearizável~\cite{linearizability}, i.e., verifica-se que, se todas operações que foram efetuadas com sucesso durante a execução ocorreram atomicamente, segundo a ordem de tempo real, e.g. se uma operação A for concluída antes do início de uma operação B, então a execução de A deve sempre preceder à execução de B, ou seja, a execução de B deve sempre observar os efeitos da execução de A.
O JepREST pode ser visto como um sistema que permite simplificar a utilização do Jepsen\cite{jepsenwebsite}, uma biblioteca construída para analisar a correção de sistemas distribuídos, no teste de aplicações REST. 
No desenho do JepREST foi necessário endereçar múltiplos desafios, incluindo 
como construir automaticamente o código necessário para a execução dos testes a partir da especificação da interface REST, incluindo a criação de novos objetos e como definir o modelo necessário à verificação da correção da execução.

O JepREST está ainda em desenvolvimento, com a versão atual suportando apenas o teste de aplicações a correrem num único servidor onde as operações seguem a semântica \textit{standard} do REST. As experiências preliminares 
efetuadas permitiram verificar que o sistema consegue detetar um 
conjunto alargado de violações da correção. 

O resto do artigo,
a secção~\ref{related_work} discute o trabalho relacionado;
a secção~\ref{background} introduz as bibliotecas usadas pelo JepREST;
a secção~\ref{system} descreve o sistema;
a secção~\ref{eval} apresenta os testes efetuados e a 
secção~\ref{conclusion} conclui o artigo.

\section{Trabalho Relacionado}\label{related_work}

A verificação da correção de um sistema distribuído é um processo complexo, devido a vários fatores, entre os quais
o funcionamento descentralizado do sistema, com os componentes a executarem ações independentes, 
a concorrência das ações executadas em diferentes componentes e em cada componente e as falhas parciais que podem 
ocorrer nas comunicações e nos próprios componentes. 
Nos testes de \textit{software} é possível decompor a verificação num conjunto de etapas~\cite{nidhra2012black}. 
Os testes unitários pretendem verificar a correção de cada componente isoladamente.
Os testes de integração pretendem verificar a correção da integração de múltiplos componentes distribuídos, podendo-se 
considerar componentes a executar na mesma máquina física ou em diferentes máquinas.
Os testes funcionais pretendem verificar que o sistema como um todo funciona de forma correta, i.e., de acordo com a especificação. 

Para que os testes funcionais dum sistema distribuído sejam significativos é necessário considerar vários fatores.
Por um lado, simular cargas de trabalho que correspondam às utilizações esperadas e a casos limites
de utilização do sistema. Por outro lado, é necessário considerar o modelo de falhas definido para o sistema,
i.e., quais as falhas que o sistema deve conseguir tolerar, simulando-as de forma a verificar que o sistema 
consegue efetivamente tolerar estas falhas.  

Múltiplos trabalhos abordam a problemática da introdução de falhas em sistemas distribuídos, incluindo a injeção de falhas nos componentes~\cite{faultsee,Natella16Assessing,Madeira00Emulation} e nas 
comunicações~\cite{nftape}. Estes trabalhos são complementares ao nosso e poderiam ser utilizados para simular falhas durante a execução dos testes.

Existem igualmente vários sistemas para a execução de testes de carga~\cite{jiang2008automatic}, que permitem avaliar 
propriedades não funcionais, como latência, desempenho e escalabilidade de aplicações distribuídas. 

\textbf{Testes de APIs REST}
Existem múltiplas ferramentas e bibliotecas que podem ser usadas para testar serviços com APIs REST.
O cURL~\cite{curl} fornece uma biblioteca que permite efetuar chamadas que sejam feitas sobre HTTP e pode ser utilizado na construção de aplicações de teste. Aplicações como o Postman~\cite{postman} e o SoapUI~\cite{soapui} 
permitem criar \textit{scripts} para testar APIs REST, definindo a sequência de operações que devem ser invocadas, assim como o resultado esperado. No entanto, nenhuma destas ferramentas permite efetuar pedidos concorrentes. 

O Dredd~\cite{dredd} permite gerar testes a partir de uma descrição da API, no formato \textit{API Blueprint} ou \textit{Swagger},
simplificando o processo de teste. No entanto, a análise dos resultados limita-se a confirmar se os mesmos estão de acordo com o formato esperado, não lidando com possíveis pedidos concorrentes nem verificando a correção do conteúdo da resposta.

O ReqBin~\cite{reqbin} é uma ferramenta \textit{online} que permite testar serviços REST e suporta a execução de pedidos concorrentes a partir de múltiplas localizações geográficas. Contudo, não suporta a análise da \mbox{correção dos resultados obtidos.} 

Existem várias ferramentas que permitem efetuar testes de carga a serviços REST~\cite{JMeter,Locust,Taurus,artillery-docs}.
Nestas aplicações os programadores especificam os testes a efetuar, indicando a sequência de operações que devem ser executadas e
a frequência relativa de cada sequência. As ferramentas executam os testes, simulando múltiplos clientes concorrentes que
executam as várias operações definidas. Em geral, os programadores podem definir qual o resultado esperado depois duma invocação, 
mas esta funcionalidade é difícil de utilizar numa situação em que o resultado dependa de operações executadas anteriormente
e que possa ser influenciado por eventuais operações concorrentes.
O nosso trabalho difere destas ferramentas em vários pontos, em particular no suporte à verificação da correção dos 
resultados, o qual é feito usando a biblioteca Knossos~\cite{jepsenKnossosGitHub} e na possibilidade de introduzir vários tipos de falhas nos vários componentes da aplicação.

Outros trabalhos abordaram a verificação da correção de serviços Web existentes~\cite{Freitas16Characterizing} e de bases 
de dados~\cite{cobra}. Ao contrário destes, o nosso trabalho foca-se no teste de sistemas distribuídos genéricos que apresentam interfaces REST.

\section{Contexto}\label{background}

Nesta secção apresenta-se o contexto em que o trabalho foi realizado, incluindo as bibliotecas usadas na construção do JepREST.


\noindent 
\textbf{Especificação OpenAPI}
(OAS)~\cite{openAPI} define uma forma de especificar um serviço web RESTful de forma independente da linguagem utilizada para desenvolver o serviço.
Esta especificação permite, usando um conjunto de anotações,
indicar quais as operações do serviço, quais os parâmetros e resultados destas operações, incluindo o esquema dos objetos manipulados e ligações entre atributos de diferentes objetos, entre outras informações.

Existem ferramentas que permitem transformar uma especificação OpenAPI em especificações utilizadas em diferentes linguagens, como o JAX-RS usado em Java, e vice-versa, simplificando a geração da especificação OpenAPI a partir de uma implementação existente.

\noindent 
\textbf{PETIT}
O PETIT~\cite{tese} é uma ferramenta de teste de microserviços, que utiliza uma especificação OpenAPI estendida com a possibilidade de definir invariantes sobre os dados e pré e pós-condições
para a execução das operações, usando lógica de primeira ordem. 
Para além de realizar pedidos à API dos serviços e avaliar os resultados obtidos, o PETIT permite gerar dados relevantes para os testes, de acordo com os esquemas definidos no documento da especificação.

\noindent 
\textbf{Jepsen}
\label{related_work:jepsen}
O Jepsen \cite{jepsenwebsite} é uma biblioteca em clojure utilizada na criação de programas para testar a correção de sistemas distribuídos, 
em particular de bases de dados.
O teste de uma base de dados distribuída usando o Jepsen é feito em ambiente Docker, com o programa de teste a executar num \textit{container} e a base de dados a executar em múltiplos outros \textit{containers}.

O programa de teste executa diversos clientes da base de dados concorrentemente, os quais executam a sequência de 
operações indicadas pelo \textit{thread} que coordena os testes. Durante o teste, podem ser injetadas no sistema um conjunto de 
falhas, sendo que o conjunto de operações e o conjunto de falhas são ambos definidos pelo programador.

Durante a execução de um teste é armazenada a história~\cite{consistency} da execução, que inclui as operações executadas e o resultado obtido. O Jepsen fornece duas ferramentas, o Knossos e o Elle~\cite{jepsenElleGitHub}, para verificar a correção da execução.
O Knossos verifica se uma história é linearizável com base num modelo do sistema, que inclui a definição do 
comportamento das operações e do estado do sistema. Enquanto o Knossos analisa as operações de forma individual, 
o Elle permite estender a validação da correção da história à execução de transações compostas por múltiplas operações.

\textbf{Faker} 
O Faker~\cite{faker} é uma biblioteca que permite gerar dados realistas, que podem ser utilizados no processo de teste, em grande escala, de aplicações com interfaces REST. 
A vantagem da utilização desta biblioteca é a sua capacidade de gerar dados relevantes, ou seja, dados semelhantes aos dados que seriam submetidos por utilizadores reais. Por exemplo, caso a aplicação necessite de um valor que seja um nome de um utilizador, a biblioteca Faker vai gerar um nome real, i.e., um nome possível para um utilizador real.

\section{JepREST}
\label{system}

O JepREST é um sistema desenvolvido para verificar a 
correção de aplicações com interfaces REST, após a submissão de um teste funcional.
Para isso, o sistema JepREST recorre à utilização 
da biblioteca Jepsen, apresentada
na secção~\ref{related_work:jepsen}, para gerar e submeter um 
conjunto de testes funcionais com múltiplos clientes a efetuarem 
operações concorrentes à aplicação REST em análise. 
De seguida, o JepREST analisa as respostas
que foram geradas pela aplicação, de forma a verificar 
se as mesmas correspondem a um comportamento correto.

\begin{figure}[t]
  \centering
  \includegraphics[height=3cm]{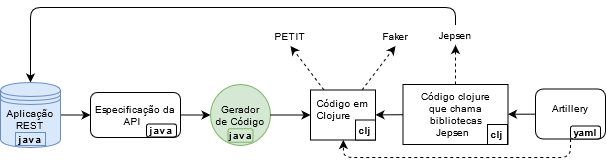}
\vspace{-0.4cm}
  \caption{Passos do teste de aplicações usando o JepREST.}
  \label{fig:API}
\vspace{-0.4cm}
\end{figure}

O teste de aplicações utilizando o JepREST consiste na seguinte sequência de passos, demonstrados na figura~\ref{fig:API}:

\begin{enumerate}[nosep]
    \item Definição da API da aplicação em estudo, assim
     como, da semântica associada às operações. A partir da API é gerado o código necessário à invocação dos métodos do serviço, incluindo a criação de novos objetos. 
    \item Definição do conjunto de testes a serem executados
     pela sistema de teste.
     O JepRest gera código para executar as cargas de trabalho definidas num \textit{script} YAML.
    \item Execução das cargas de trabalho, utilizando o código gerado anteriormente, recorrendo às bibliotecas do Jepsen para controlar a execução dos testes e obter o \textit{log} da execução.
    \item Análise dos resultados obtidos, verificando se a execução corresponde a um comportamento correto, definido como uma execução que seja linearizável segundo o modelo da API definido.
\end{enumerate}

De seguida detalha-se cada um dos passos da solução.
\subsection{Definição da API da aplicação}
\label{jeprest:api}

Como o JepREST vai necessitar de utilizar os métodos e as estruturas oferecidas pelo Jepsen, que estão desenvolvidas em clojure, foi necessário criar um gerador de código para processar a especificação da aplicação e gerar métodos em clojure que vão ser executados pelo Jepsen para fazer as chamadas necessárias à aplicação REST em estudo. A especificação do serviço REST deve ser fornecida pelo programador em código Java, com anotações do OpenAPI e do JAX-RS.

A biblioteca Jepsen oferece clientes que efetuam operações de leitura e de escrita simples, contudo não contém
nenhum cliente que permita submeter operações REST. Por isso, foi necessário criar um novo cliente Jepsen e definir para cada tipo de operação REST o comportamento do cliente para o pedido ser executado. 
Para cada operação definida na API, define-se o comportamento da mesma através de um método clojure gerado que é executado por um cliente quando for submetido um pedido da operação correspondente. Para a geração destes métodos, assume-se que a aplicação em estudo adota a semântica REST \textit{standard}, em que: o POST consiste na criação de novos recursos; o GET consiste na leitura de um ou vários recursos; o PUT consiste em atualizar um recurso específico; o PATCH consiste em atualizar parcialmente um recurso já existente e, por fim, o DELETE consiste na remoção de um recurso específico.

\subsection{Definição do conjunto de testes}

Para determinar a correção de uma aplicação REST é necessário analisar
as respostas do sistema quando é executado um conjunto de testes significativo.
Existem dois desafios neste processo. Primeiro, garantir que as operações 
são invocadas com parâmetros relevantes. Segundo, definir cargas de trabalho
que sejam apropriadas para o sistema.
    
Relativamente ao primeiro desafio, se fosse definido um conjunto de operações onde apenas são emitidas operações GET, PUT, PATCH ou DELETE de recursos
que nunca foram criados, os resultados gerados pela aplicação 
não serão muito informativos sobre a implementação da 
aplicação em estudo.
Para lidar com este problema, o código gerado mantém informação nos clientes de quais os objetos criados nos servidores e utiliza a anotação @Link do OpenAPI para estabelecer ligações entre diferentes objetos. Estas anotações definem dependências entre as diferentes operações, e.g. se uma operação A apresenta uma anotação @Link para um operação B, como no exemplo da figura~\ref{fig:POST_Link} considera-se que a operação B é dependente da operação A, uma vez que a mesma para ser efetuada com sucesso necessita de receber como parâmetro a informação da operação~A.

\begin{figure}[t]
  \centering
  \begin{lstlisting}[language= Java, frame=single, basicstyle=\scriptsize]
@POST
@Operation(operationId = "createStudent", responses = {
    @ApiResponse(links = {@Link(name = "GetStudentByID", 
    operationId = "getStudent", 
    parameters = {@LinkParameter(name = "studentId", 
                   expression = "$response.body#/id")})})})
Response createStudent(Student student);
  \end{lstlisting}
\vspace{-0.4cm}
  \caption{Exemplo de uma anotação @Link criada.}
  \label{fig:POST_Link}
  \vspace{-0.4cm}
\end{figure}

Após o processamento das anotações JAX-RS de cada operação da especificação para gerar os métodos clojure que descrevem o comportamento de um cliente quando se deseja que essa operação seja submetida à aplicação, o gerador de código vai analisar e processar todas as anotações @Link existentes na especificação da aplicação, para detetar todas as dependências entre as operações e armazenar em listas toda a informação de uma operação específica que possa ser utilizada como \textit{input} noutras operações da aplicação.
De seguida, o gerador de código é responsável por gerar os métodos clojure que vão definir as invocações que serão feitas à aplicação REST. Para isso, o gerador vai utilizar as listas que criou para selecionar os valores dos \textit{inputs} das operações que são dependentes de outras, e.g. parâmetros.

Um segundo aspeto relativo a definir invocações com parâmetro relevantes prende-se com a geração de novos objetos - e.g. a informação dum estudante. 
Para endereçar este desafio, o JepREST recorre às bibliotecas Faker e PETIT, para a geração automática dos dados. O programador deve definir na especificação da API como os diferentes dados devem ser gerados, recorrendo ao uso de diferentes tipos de anotações, a figura~\ref{fig:Dados_Annot} é um exemplo de como os dados podem ser especificados. 
Caso exista a anotação especial @Values, o gerador de código define que esse dado vai ser gerado pela ferramenta Faker, onde será evocado o método descrito na anotação. Caso sejam utilizadas anotações @Min, @Max, @Size ou @Pattern sem a anotação @Values, o gerador de código, por sua vez, define que esse dado será gerado pela ferramenta PETIT, onde os valores das anotações serão enviados como parâmetros dos métodos do PETIT que vão ser chamados. 

Desta forma, irá se utilizar sempre que possível a ferramenta Faker, uma vez que a mesma gera dados realistas, que poderiam ser fornecidos durante a utilização da aplicação por um utilizador real.


\begin{figure}[t]
  \centering
  \begin{lstlisting}[language=Java, frame=single, basicstyle=\scriptsize]
    @Values("faker.name/first-name") @Pattern(regexp = "[A-Z][a-z]+")
    private String firstName;
    @Min(0) @Max(100)
    private int age;
    @Pattern(regexp = "[A-Z][a-z]+") @Size(min = 20, max = 500)
    private String description; 
  \end{lstlisting}
 \vspace{-0.4cm}
 \caption{Exemplo de anotações que representam as propriedades dos dados.}
  \label{fig:Dados_Annot}
   \vspace{-0.4cm}
\end{figure}
 
O segundo desafio prende-se com a definição das cargas de trabalho 
a serem submetidas à aplicação durante o processo de teste.
No JepREST, as cargas de trabalho são definidas num \textit{script} YAML inspirado na ferramenta Artillery\footnote{É de realçar que ao contrário
dum \textit{script} de testes Artillery, que especifica os detalhes da invocação das operações REST (incluindo entre outros o método HTTP, os parâmetros e a sua codificação), o \textit{script} usado pelo JepREST
apenas indica o nome das operações.},
especificando diferentes cenários de teste, compostos por sequências de operações, definidas na \textit{tag flow}, (definidas usando o nome do método clojure gerado no passo anterior) a serem executadas concorrentemente por vários clientes, onde cada cenário terá um dado peso. Os pesos são relevantes para determinar qual o cenário que deve ser executado um maior número de vezes, e.g. se num ficheiro YAML existir o cenário 1 com peso 90 e o cenário 2 com peso 10, a sequência de operações definida no primeiro cenário executará 9 vezes mais frequentemente que a sequência de operações definidas no segundo cenário. 




\subsection{Execução dos testes}

Antes do JepREST submeter um teste à aplicação em estudo, o programador é responsável por definir a localização dos \textit{containers} onde os servidores REST existentes estão a correr, i.e. endereço IP e a porta, para que os clientes possam saber aonde é que podem submeter os seus pedidos.

Assume-se que sempre que um teste é executado, a aplicação é inicializada, o JepREST não suporta a verificação de aplicações já em funcionamento, onde já foram efetuados pedidos REST. Portanto, antes da execução de um teste o serviço é inicializado e lançado em ambiente Docker - a aplicação que se está a testar está a correr em um ou mais \textit{containers Docker}. O JepREST permite verificar a correção tanto se estes \textit{containers} estiverem a correr localmente ou remotamente, apenas é necessário informar a localização dos mesmos.
Quando o JepREST executa um teste, o mesmo é responsável por gerar e enviar ao Jepsen a sequência de operações que foi definida. De seguida, o \textit{container} de controlo do Jepsen vai criar vários clientes e informá-los sobre as operações que os mesmos devem submeter à aplicação, assim como toda a informação necessária dos pedidos, inclusive a localização do \textit{container}, onde o servidor a qual se pretende fazer o pedido está a correr.

Algumas funcionalidades ainda em implementação neste componente do sistema incluem a definição do conjunto de 
falhas a simular e o suporte de múltiplos e distintos serviços REST para a mesma aplicação. 

\subsection{Validação dos resultados}

Após a conclusão de um teste, o JepREST vai analisar as suas respostas para 
verificar se as mesmas correspondem a um funcionamento correto. 
A informação da história da execução é armazenada num ficheiro 
automaticamente gerado pelo Jepsen, durante a execução das 
várias operações.

O JepREST analisa a história de operações efetuadas recorrendo
ao \textit{checker} Knossos, oferecido pelo Jepsen. 
Este \textit{checker} é responsável por verificar se a 
história de operações é linearizável segundo o modelo que representa
o estado e como o mesmo deve ser modificado na presença de operações.

Para usar o Knossos, o JepREST define e implementa o modelo
do estado da aplicação, onde é expressa a forma como o mesmo é alterado na presença de uma operação REST específica.

A implementação do modelo foi dividida em duas fases, a primeira onde foi escolhido um tipo de dados para representar o estado da aplicação REST em estudo, e a segunda onde se determina qual deve ser a mudança do estado da aplicação após a execução de uma operação REST específica.
Na primeira fase, definiu-se que o estado de uma aplicação REST iria ser representado através de um mapa, onde as chaves são o nome dos recursos e os valores são outros mapas em que as chaves são os identificadores dos recursos que já foram criados e os valores são os objetos \textit{json} que representam os recursos existentes. 

Na segunda fase é definido o novo estado da aplicação, em resultado da aplicação das várias operações REST definidas.
Por exemplo, um POST devolve um erro caso o identificador do objeto já exista e caso não exista atualiza o estado da aplicação com
o objeto recebido. As restantes operações são definidas de forma semelhante, considerando as situações de erro e as atualizações 
correspondentes.
\section{Experiências}
\label{eval}

\begin{figure}[t]
  \centering
  \includegraphics[height=3cm]{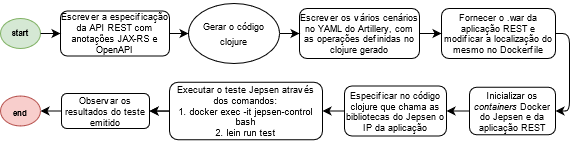}
\vspace{-0.4cm}
  \caption{Ações de um utilizador para executar o JepREST.}
  \label{fig:Actions_JepREST}
  \vspace{-0.3cm}
\end{figure}

A figura~\ref{fig:Actions_JepREST} mostra os passos que um utilizador tem de seguir para utilizar a ferramenta JepREST para testar uma aplicação com interface REST. O maior esforço de um programador que pretende utilizar o JepREST é ter de escrever a especificação da API em estudo, indicar a localização do servidor REST, providenciar o ficheiro .war da API e indicar a sua localização no Dockerfile da API. Os restantes passos são praticamente automáticos onde apenas é necessário executar comandos no terminal ou mesmo executar código, como por exemplo, correr o código do gerador ou o código Clojure para submeter testes à aplicação.

Para verificar se o sistema JepREST avalia corretamente as execuções de aplicações REST quando são emitidos pedidos concorrentes, foi desenvolvida uma aplicação, escrita em Java, onde é utilizada uma base de dados H2 \textit{embedded} e foi criado apenas um recurso e 6 \textit{endpoints} diferentes: criar um estudante (POST), devolver um estudante com o dado identificador (GET), devolver todos os estudantes (GET), modificar um estudante com o identificador dado (PUT), modificar parcialmente um estudante com o identificador dado (PATCH) e remover um estudante com o identificador dado (DELETE). Ao implementar cada um destes \textit{endpoints} optou-se por não utilizar transações sempre que fosse necessário fazer modificações à base de dados. Tomou-se esta decisão uma vez que existirão problemas de consistência num ambiente de concorrência e pretende-se confirmar que o JepREST consegue detetar estes problemas nas várias execuções da API, onde são emitidos os vários tipos de pedidos REST.

Após a implementação da API, a escrita da sua especificação e a geração do código clojure, escreveu-se o teste a correr no ficheiro YAML, onde se definiu um cenário para ser executado várias vezes a cada 1 milissegundo, durante 30 segundos, demonstrado na figura~\ref{fig:cenario}. Este cenário contêm uma sequência de operações que vão ser submetidas à aplicação por vários clientes, de forma concorrente, mesmo se ocorrerem falhas a meio da sequência, as operações continuam a ser emitidas pelos vários clientes do JepREST.

\begin{figure}[t]
  \centering
  \begin{lstlisting}[frame=single, basicstyle=\scriptsize]
scenarios:
    weight: 100
    flow:
      - createStudentData
      - updateStudentData
      - getAllStudentsData
      - getStudentData
      - deleteStudentData
  \end{lstlisting}
\vspace{-0.4cm}
  \caption{Cenário executado pelo sistema JepREST.}
  \label{fig:cenario}
\end{figure}

O JepREST submeteu este teste, várias vezes à aplicação criada, e detetou diferentes problemas de consistência em quase todas as execuções.
De seguida, são apresentados os resultados das várias execuções do teste definido, com uma breve descrição sobre os problemas de consistência que foram detetados.

\noindent \textbf{Experiência 1.}
A figura \ref{fig:error_get1} demonstra parte de um \textit{log} gerado pelo JepREST após a execução do teste definido, onde são retratadas as sequências de operações que foram submetidas assim como as respostas da aplicação após a execução das mesmas. Uma linha do \textit{log} tem a seguinte informação, respetivamente da esquerda para a direita: um \textit{timestamp} que representa o instante que a operação foi invocada/executada, o identificador do cliente Jepsen responsável por submeter/receber a operação, uma indicação sobre se a operação é uma invocação (invoke) ou uma resposta (ok/fail), o tipo de operação REST (post/put/get/patch/delete) e, por fim, informação extra sobre a operação, como o \textit{input} e o \textit{output}.

\begin{figure}[t]
  \centering
  \begin{lstlisting}[numbers = left, frame=single,basicstyle=\scriptsize]
11:59:59 :4 :invoke, :put, {:firstName "Brycen", :lastName "Cummerata",
            :email "adam@prince.com", :age 129, :phone "119-364-8408"},
            :path "498C98D9E8CB"
11:59:59 :0 :invoke, :get
11:59:59 :2 :invoke, :get, :path "498C98D9E8CB"
11:59:59 :3 :invoke, :delete, :path "498C98D9E8CB"
11:59:59 :4 :ok, :put, :output {:id "498C98D9E8CB", 
            :firstName "Brycen",:lastName "Cummerata", 
            :email "adam@prince.com", :age 129, :phone "119-364-8408"}
11:59:59 :3 :ok, :delete, :output "498C98D9E8CB"
11:59:59 :0 :fail, :get
11:59:59 :2 :ok, :get, :path "498C98D9E8CB", :output {:id "498C98D9E8CB", 
            :firstName "Brycen",:lastName "Cummerata", 
            :email "adam@prince.com", :age 129, :phone "119-364-8408"}
  \end{lstlisting}
\vspace{-0.4cm}
  \caption{Parte de um \textit{log} gerado pelo JepRest durante uma execução do teste definido.}
  \label{fig:error_get1}
  \vspace{-0.4cm}
\end{figure}

Este \textit{log} retrata nas linhas 1, 4, 5 e 6, a submissão de quatro operações REST distintas à aplicação, de forma concorrente e ao observar as respostas da aplicação, destes pedidos, verifica-se que na linha 10 a operação DELETE é efetuada com sucesso, ou seja, o estudante com o identificador ``498C98D9E8CB'' é removido, contudo, na linha 12, a aplicação responde que a operação GET é efetuada com sucesso e o estudante com o identificador ``498C98D9E8CB'' é recebido. Isto acontece, porque o estudante com identificador dado é removido depois de ser verificado se o mesmo existe pela operação GET. Contudo, como a operação DELETE foi efetuada com sucesso (linha 10) antes da operação GET (linha 12), o JepREST reporta um erro de consistência, afirmando que a operação GET não pode ter sido efetuada com sucesso, visto que o identificador do estudante dado já foi removido.

\noindent \textbf{Experiência 2.}
Esta experiência retrata outra execução do mesmo teste à aplicação REST implementada, onde são obtidos resultados diferentes da primeira experiência. Parte do \textit{log} gerado pelo JepRest está apresentado na figura~\ref{fig:error_get2}, e este apresenta a mesma informação que o \textit{log} da experiência anterior.  

\begin{figure}[t]
  \centering
      \begin{lstlisting}[numbers = left, frame=single,basicstyle=\scriptsize]
12:30:17 :2 :invoke, :put, {:firstName "Sasha", :lastName "Hyatt", 
         :email "claudine.prosacco@hotmail.com", :age 93, 
         :phone "(165)479-5262 x15024"}, :path "71D1083D76BD"
12:30:17 :3 :invoke, :get
12:30:17 :1 :invoke, :get, :path "71D1083D76BD"
12:30:17 :0 :invoke, :delete, :path "71D1083D76BD"
12:30:17 :1 :ok, :get, :path "71D1083D76BD", :output {:id "71D1083D76BD",
         :firstName "Grayce", :lastName "Brekke",
         :email "marques.rodriguez@yahoo.com", :age 40, 
         :phone "1-162-508-6862 x13530"}
12:30:17 :3 :ok, :get, :output ({:id "71D1083D76BD", 
         :firstName "Grayce", :lastName "Brekke", 
         :email "marques.rodriguez@yahoo.com", :age 40, 
         :phone "1-162-508-6862 x13530"})
12:30:17 :0 :ok, :delete, :output "71D1083D76BD"
12:30:17 :2 :ok, :put, :path "71D1083D76BD", :output nil
  \end{lstlisting}
\vspace{-0.4cm}
  \caption{Parte de um \textit{log} gerado pelo JepRest durante uma execução do teste definido.}
  \label{fig:error_get2}
\vspace{-0.4cm}
\end{figure}
Tal como no \textit{log} da Experiência 1, pode-se observar nas linhas 1, 4, 5 e 6 vários pedidos concorrentes de diferentes clientes à aplicação REST. Nesta experiência, o JepREST deteta um erro de consistência na linha 16, uma vez que é observado que a operação responsável por atualizar o estudante com o identificador "71D1083D76BD" é efetuada com sucesso, contudo a operação responsável por remover o estudante com este identificador foi efetuada com sucesso momentos antes, na linha 15. O sistema criado, JepREST, deteta esta inconsistência e reporta-a informando que o resultado da operação PUT está incorreto (linha~15), visto que o identificador do estudante que pretende atualizar já foi removido pela operação DELETE (linha 16).

\section{Conclusões e Trabalho Futuro}
\label{conclusion}

O sistema JepREST permite simplificar o teste de aplicações distribuídas com interfaces REST, assim como, a verificação da sua correção.
A versão atual do sistema inclui todos os componentes necessários ao teste de aplicações que adotam o modelo REST \textit{standard}.
As experiências realizadas mostram que o JepREST permite detetar problemas de correção das aplicações que apenas surgem 
devido à concorrência entre múltiplos clientes. 

%
%

O sistema JepREST encontra-se ainda em desenvolvimento, em particular as seguintes funcionalidades.
Primeiro, o suporte à execução das aplicações está a ser estendido para suportar a injeção de falhas durante a execução.
As falhas a introduzir serão especificadas pelos programadores num ficheiro de \textit{script}.
Segundo, pretende-se suportar operações que implementam diferentes semânticas, incluindo a possibilidade de definir transações.
Para tal, será necessário usar o Elle~\cite{jepsenElleGitHub} para verificar a correção do aplicação.


%
%
{\footnotesize
\vspace{-0.2cm}
\paragraph*{Agradecimentos:}
Agradecemos a Kyle Kingsbury pelos comentários ao sistema JepREST.
Este trabalho foi parcialmente financiado pelos projectos 
UIDB/04516/2020, 
PTDC/CCI-INF/32662/2017 e PTDC/CCI-INF/32081/2017 da FCT/MCTES.
}

\end{document}